\documentclass[12pt,a4paper]{article}
\usepackage{setspace}
\usepackage[english]{babel}
\usepackage{amssymb}
\usepackage{amsmath,amsfonts}
\usepackage{txfonts}
\usepackage{mathdots}
\usepackage[classicReIm]{kpfonts}
\usepackage{graphicx}
\usepackage[utf8]{inputenc}
\usepackage[T1]{fontenc}
\usepackage{amsfonts}
\usepackage{makeidx}
\usepackage{ragged2e}
\usepackage[labelfont=bf]{caption}
\usepackage{subcaption}
\date{}
\usepackage[numbers]{natbib}
\usepackage[left=2.54cm, right=2.54cm, top=2.54cm, bottom=2.54cm]{geometry}

\author{S. D. Katore$^1$, S. P. Hatkar$^2$, D. P. Tadas$^3$\footnote{Corresponding author E-mail: dtadas144@rediffmail.com}}

\title{Accelerating Kaluza-Klein Universe in Modified theory of gravitation}

\begin{document}
\bibliographystyle{abbrvat}
\maketitle

\begin{center}
\onehalfspacing
	$^1$Department of Mathematics, Sant Gadge Baba Amravati University\\ Amravati-444602, India.\\
	\small{E-mail: katore777@gmail.com}\\
	$^2$Department of Mathematics, A. E. S. Arts, Commerce and Science\\ College, Hingoli-431513, India.\\
	\small{E-mail: schnhatkar@gmail.com}\\
	$^{3*}$Department of Mathematics, Toshniwal Arts, Commerce and Science,\\ College Sengaon-431542, India.\\
	\small{E-mail: dtadas144@rediffmail.com}\\
\end{center}
\doublespacing
\begin{abstract}

\noindent The purpose of this paper is to study the Kaluza-Klein universe in the context of the $ f(R,T) $ gravity theory using magnetized strange quark matter (MSQM). To obtain exact solutions of field equations, we assume two types of volumetric expansion: power law and exponential law volumetric expansions. The violation of energy conditions has been studied. The physical and geometrical properties of the examined model have also been investigated thoroughly.

\end{abstract}

\section{Introduction:}
The study of cosmic accelerated expansion, which has been validated by various observations over the last two decades, is one of the most important cosmological egnima among cosmologists. The concept of accelerated expansion of the universe was first proposed by cosmological studies such as type Ia supernovae \cite{Riess et al. 1998,Riess et al. 1999,Perlmutter et al. 1999,Astier et al. 2006}, and subsequently, measurements of the cosmic microwave background (CMB) from the wilkinson microwave anisotropy probe (WMAP) \cite{Bennett et al. 2003,Spergel et al. 2003} and large-scale structure \cite{Tegmark et al. 2004} have confirmed this idea. There are several modified theories proposed that can be found in the literature to explain the accelerating and expanding nature of the universe. Some of theme are $ f(R)$ theory \cite{Buchadahl et al. 1970,Capozziello et al. 2006,Nojiri et al. 2017},  $ f(T)$ theory \cite{Ferraro et al. 2007,Farajollahi et al. 2012,Rodrigues et al. 2013}, $ f(R,T)$ theory \cite{Harko et al. 2011,Houndjo et al. 2012}, $f(G)$ theory \cite{Nojiri and Odintsov 2005,Cognola et al. 2006,Nojiri and Odintsov 2011} and each theory has its own importance.

Among the several modified theories of gravitation, the $f(R,T)$ theory of gravity proposed by Harko et al.\cite{Harko et al. 2011} is an intriguing extension of general relativity (GR) that has received a lot of attention in recent years. The late time cosmic accelerated expansion of the universe can be explained by the $ f(R,T) $ gravity theory. Houndjo et al. \cite{Houndjo et al. 2012} used an auxiliary scalar field with two known forms of scale factor to reconstruct $f(R,T)=f(R)+f(T)$ and obtained a transition from a matter-dominated phase to an accelerated phase. Sharif and Zubair \cite{Sharif and Zubair 2012} have studied the law of thermodynamics in $f(R,T)$ gravity theory. Using bulk viscous fluid, Chandel et al. \cite{Chandel et al. 2012} studied hypersurface homogeneous cosmological models with time-dependent cosmological terms.  In the presence of the perfect fluid source, Reddy and Santhi \cite{Reddy and Santhi 2013} investigated LRS Bianchi-II space time. The LRS Bianchi-I bulk viscous cosmological models in $f(R,T)$ gravity have been investigated by Mahanta et al. \cite{Mahanta et al. 2014}. Moreover, Harko and Lake \cite{Harko and Lake 2015} have investigated Kasner-type static, cylindrically symmetric interior string solutions using the $f(R,L_{m} )$ theory of gravity. Pawar and Solanke \cite{Pawar and Solanke 2015} have explored prefect fluid LRS Bianchi-I cosmological models in $f(R,T)$ theory of gravity. In the context of $f(R,T)$ gravity theory, Singh and Singh \cite{Singh and Singh 2015} discussed the behaviour of a flat FRW cosmological model with a scalar field. Mishra et al. \cite{Mishra et al. 2016} have studied perfect fluid Bianchi-VI${}_{h}$ space time in $f(R,T)$ theory of gravity.

The Kaluza-Klein theory is a unification of Einstein’s theory of gravitation and Maxwell's theory of electromagnetism by introducing compactified extra dimensions. It is regarded as a crucial forerunner to string theory and has received a lot of interest in recent years. Kaluza and Klein explained the role of electromagnetic field in the fremework of a five-dimensional space time. Recently, Mishra et al. \cite{Mishra et al. 2019} compared the Kaluza-Klein dark energy in the Lyra manifold with general relativity using magnetic field. Aktas \cite{Aktas 2019} explored the behaviour of Kaluza-Klein massive and massless scalar field cosmological models with $\Lambda$ in $f(R,T)$ gravity theory. Hatkar and Katore \cite{Hatkar and Katore 2020} have used Polytropic Equation of State in Lyra Geometry to examine the Kaluza-Klein space time.

Strange quark matter (SQM) is a new kind of matter made up of many deconfined up ($ u $), down ($ d $), and strange ($ s $) quarks \cite{Bodmer 1971,Witten 1984} and its properties are studied in equilibrium with the weak interactions \cite{Farhi and Jaffe 1984} using the MIT bag model \cite{Chodos et al. 1974}. Furthermore, in the context of the MIT bag model, the thermodynamical properties of SQM were examined in a strong magnetic field with quark confinement by density dependence quark masses, taking total baryon density, charge neutrality, and $ \beta $-equilibrium at zero temperature into account\cite{Felipe et al. 2008,Isayev and Yang 2012,Jia-Xun et al. 2015}. Chakrabarty \cite{Chakrabarty 1996} used the conventional MIT bag model to investigate the effect of a strong magnetic field on the stability and properties of SQM. Singh and Beesham \cite{Singh and Beesham 2021} have examined the LRS Bianchi-I cosmological model in $ f(R,T) $ gravity using SQM. Also, Magnetism influences the anisotropies in CMB radiation, which also plays an important role in the formation of structures. Magnetic fields have been observed in the high redshift Lyman system, galaxies, clusters and stars. One of the most interesting areas of research is the relation between magnetic fields and SQM. The behaviour of magnetized strange quark matter (MSQM) for LRS Bianchi-I model has studied in $f(R,T)$ gravity by Sahoo et al. \cite{Sahoo et al. 2017}. Aktas \cite{Aktas 2017} have studied Bianchi I and V models with MSQM distributions in reconstructed $f(R,T)$ theory of gravity. Aktas and Aygun \cite{Aktas and Aygun 2017} have investigated FRW space time with MSQM solutions in $f(R,T)$ theory of gravity. Also, some authors like \cite{Miransky and Shovkovy 2015,Isayev and Yang 2013,Li et al. 2016,Tsagas and Borrow 1997} have studied various aspects of MSQM cosmological models in $f(R,T)$ theory of gravitation. Recently, Khalafi and Malekolkalami \cite{Khalafi and Malekolkalami 2021} has studied the MSQM Bianchi-I space time for two different models of $f(R,T)$ thoery.

In the present research work, we investigate MSQM distributions for the Kaluza-Klein metric in the context of $f(R,T)$ gravity theory, which is motivated by the discussion above and a few MSQM studies in the literature. Section 2 is devoted to the metric and $f(R,T)$ gravity field equations for $f(R,T)=R+2f_{1} (T)$. The solutions to the field equation for the power law and exponential volumetric expansion models are derived in sections 3 and 4, respectively. The energy condition and its interpretation of the investigated model are covered in Section 5. Finally, the last section is devoted to the conclusion.

\section{Metric and $f(R,T)$ gravity field equations:}
In the Kaluza-Klein model, the additional dimension's contribution to the energy momentum tensor is often due to electromagnetic field stresses. We consider the five-dimensional Kaluza-Klein space time of the form as \cite{Katore and Hatkar 2015}
\begin{equation} \label{Eq:1} 
	ds^{2} =-dt^{2} +A^{2} \left(dx^{2} +dy^{2} +dz^{2} \right)+B^{2} d\psi ^{2} , 
\end{equation}
\noindent where $ A $ and $ B $ are functions of $t$ only.
\noindent In the present study, we assume the matter contents described by energy momentum tensor for the MSQM is given in the following form as \cite{Sahoo et al. 2017,Aktas and Aygun 2017}:
\begin{equation} \label{Eq:2} 
	T_{ij} =\left(\rho +p+h^{2} \right)\, u_{i} u_{j} +\left(\frac{h^{2} }{2} -p\right)\, g_{ij} -h_{i} h_{j}  
\end{equation}
where $\rho $ is energy density, $P$ is pressure, $h^{2} $ is magnetic flux and $u_{i}=(0 ,0, 0, 0, 1) $ is the velocity vector in comoving coordinate system satisfying the condition $u_{i} u^{i} =-1$. The magnetic flux $h^{2}=h_{i}h^{i} $ is chosen in the direction of $ x-$axis satisfying $h_{i} u^{i} =0$ \cite{Barrow et al. 2007, Aktas and Yilmaz 2011}. As the flux quantizes along $ x-$axis, which gives the magnetic field in the $ yz-$plane.

\noindent The action for $f(R,T)$ gravity is expressed in the following form\cite{Harko et al. 2011}:
\begin{equation} \label{Eq:3} 
	S=\int \sqrt{-g} \left(f(R,T)+L_{m} \right)\,  d^{4} x, 
\end{equation}

\noindent where $f(R,T)$ is an arbitrary function of the Ricci scalar $R$ and the trace $T$ of the stress-energy tensor of the matter $T_{ij}$ and $L_{m} $ denotes the matter Lagrangian density.

\noindent The stress-energy tensor of the matter is defined as
\begin{equation} \label{Eq:4} 
	T_{ij} =-\frac{2}{\sqrt{-g} } \frac{(\delta \sqrt{-g}\, L_{m})  }{\delta g^{ij} }  
\end{equation}

\noindent Assuming that the Lagrangian density $L_{m}$ of the matter depends only on the metric tensor components $g_{ij} $ and not on its derivatives, we obtain
\begin{equation} \label{Eq:5} 
	T_{ij} =g_{ij} L_{m} -\frac{\delta (L_{m} )}{\delta {\rm \; }g^{ij} }
\end{equation}

\noindent The $f(R,T)$ gravity field equations are obtained by varying the action \eqref{Eq:3} with respect to the metric tensor components $g_{ij} ,$
\begin{equation} \label{Eq:6} 
	f_{R} (R,T)R_{ij} -\frac{1}{2} f(R,T)g_{ij} +\left(g_{ij} \nabla ^{i} \nabla _{i} -\nabla _{i} \nabla _{j} \right)f_{R} (R,T)=8\pi T_{ij} -f_{T} (R,T)T_{ij} -f_{T} (R,T)\Theta _{ij}  
\end{equation}

\noindent where $\nabla _{i}$  is the covariant derivative, $\Theta _{ij} =-2T_{ij} +g_{ij} L_{m} -2g^{\alpha \beta } \frac{\partial ^{2} L_{m} }{\delta g^{ij} \partial g^{\alpha \beta } } ,$ $f_{R} =\frac{\partial {\rm \; }f\left(R,T\right)}{\partial R} $ and $f_{T} =\frac{\partial {\rm \; }f\left(R,T\right)}{\partial T} $.

\noindent There are several theoretical models that can be used to represent various matter contributions to $ f(R,T) $ gravity. However, Harko et al. \cite{Harko et al. 2011} categorised these models into three distinct classes, which are as follows:
\begin{equation}\label{Eq:7}
	f(R,T)=
	\begin{cases}
		R+2f_{1} (T) \\
		f_{1} (R)+f_{2} (T) \\
		f_{1} (R)+f_{2} (R)f_{3} (T)
	\end{cases}
\end{equation}
\noindent In order to analyse the exact solutions of the Kaluza-Klein universe, we consider the first model $f(R,T)=R+2f_{1} (T)$, where $f_{1} (T)$ is an arbitrary function of the trace of the stress-energy tensor of matter. We choose the arbitrary function $f_{1} (T)$ of trace of the stress energy tensor of matter source given by
\begin{equation}\label{Eq:8}
f_{1} (T)=\mu T \Rightarrow f'_{1} (T)=\mu,
\end{equation} 
where $\mu $ is a constant and dash ( ' ) denotes differentiation with respect to the argument.

\noindent For this choice, the $f(R,T)$ gravity field equation \eqref{Eq:6} becomes
\begin{equation} \label{Eq:9} 
R_{ij} -\frac{1}{2} Rg_{ij} =(8\pi+2\mu) T_{ij} + (2\mu P+\mu T)T_{ij}
\end{equation} 

\noindent Using co-moving coordinates, we get four independent field equations of $f(R,T)$ gravity for the given metric \eqref{Eq:1} are as follows:
\begin{equation} \label{Eq:10} 
\frac{2A_{55} }{A} +\frac{B_{55} }{B} +\frac{2A_{5} B_{5} }{AB} +\frac{A_{5} {}^{2} }{A^{2} } =-\left(8\pi -3\mu \right)\frac{h^{2} }{2} -\left(8\pi +4\mu \right)p+\mu \rho  
\end{equation}
\begin{equation} \label{Eq:11} 
\frac{2A_{55} }{A} +\frac{B_{55} }{B} +\frac{2A_{5} B_{5} }{AB} +\frac{A_{5} {}^{2} }{A^{2} } =\left(8\pi +7\mu \right)\frac{h^{2} }{2} -\left(8\pi +4\mu \right)p+\mu \rho  
\end{equation}
\begin{equation} \label{Eq:12} 
\frac{3A_{55} }{A} +\frac{3A_{5} {}^{2} }{A^{2} } =\left(8\pi +7\mu \right)\frac{h^{2} }{2} -\left(8\pi +4\mu \right)p+\mu \rho  
\end{equation}
\begin{equation} \label{Eq:13} 
\frac{3A_{5} B_{5} }{AB} +\frac{3A_{5} {}^{2} }{A^{2} } =\left(24\pi +11\mu \right)\frac{h^{2} }{2} -4\mu \rho +\left(8\pi +3\mu \right)\rho  
\end{equation}
where the subscript '5' denotes differentiation with respect to t.\\
\noindent The average scale factor $ a $ and volume $ V $ of universe are defined as
\begin{equation}\label{Eq:14}
	V=a^4=A^3B
\end{equation}

\noindent Equating equations \eqref{Eq:11} and \eqref{Eq:12}, we get
\begin{equation}\label{Eq:15} 
\frac{d}{dt}\left(\frac{A_{5} }{A} -\frac{B_{5} }{B}\right) + \left(\frac{A_{5}}{A}-\frac{B_{5}}{B}\right)\frac{V_{5}}{V}=0
\end{equation}
on integrating above equation using the \eqref{Eq:14}, the values of metric potentials $A$ and $B$ are as follows
\begin{equation}\label{Eq:16} 
	A=(c_{2}V)^{1/4}exp\left[\frac{c_{1}}{4}\int \frac{dt}{V}\right]
\end{equation}
\begin{equation} \label{Eq:17} 
	B=c_{2}^{-3/4}V^{1/4}exp \left[\frac{-3c_{1}}{4}\int \frac{dt}{V}\right]
\end{equation}

Recently, Katore and Hatkar \cite{Katore and Hatkar 2015} and Sahoo et al. \cite{Sahoo et al. 2016} have investigated some interesting results of Kaluza-Klein metric using MSQM distribution. Here, we follow Sahoo et al. \cite{Sahoo et al. 2016} and Moraes \cite{Moraes 2014}.
\noindent Further, we have system of four independent equations \eqref{Eq:10} to \eqref{Eq:13} in five unknowns $ viz. $ $A,$ $B,$ $h^{2},$ $p$ and $\rho $.  We need one more condition to get the exact solutions of field equations. We consider the two different volumetric expansions such as power law expansion and exponential law expansion.
\section{Power law expansion model:}
Firstly, we consider the power law volumetric expansion as
\begin{equation}\label{Eq:18}
	V=A^{3}B=t^{4m} 
\end{equation}
where $m$ is a positive constant.

Using the equation \eqref{Eq:18}, values of $A$ and $B$ are obtained as
\begin{equation}\label{Eq:19} 
	A=c_{2}^{1/4} t^{m}exp\left[\frac{c_{1}t^{1-4m}}{4(1-4m)}\right]
\end{equation}
\begin{equation} \label{Eq:20} 
	B=c_{2}^{-3/4} t^{m}exp \left[\frac{-3c_{1}t^{1-4m}}{4(1-4m)}\right]
\end{equation}

The metric potentials A and B both vanish at time $ t = 0 $, they start to increase with time and finally diverge to infinity as  $ t \to \infty $. This is consistent with
the big bang model.
Now, the various cosmological parameters such as mean Hubble parameter $ H $, Expansion scalar $\theta $, Shear scalar $\sigma $ obtained as follows:
\begin{equation} \label{Eq:21} 
	H =\frac{1}{4} \left(\frac{3A_{5} }{A} +\frac{B_{5} }{B} \right)=\frac{m}{t}  
\end{equation}
\begin{equation} \label{Eq:22} 
	\theta=\frac{3A_{5} }{A} +\frac{B_{5} }{B} =\frac{4m}{t}  
\end{equation}
\begin{equation} \label{Eq:23} 
	\sigma ^{2}=\frac{1}{3} \left(\frac{3A_{5} }{A} -\frac{B_{5} }{B} \right)^{2} =\frac{3 c_{1}^2}{8 t^{8m}}
\end{equation}

From the expression \eqref{Eq:21}, \eqref{Eq:22} and \eqref{Eq:23}, it is observed that the Hubble parameter $ H,$ exapansin scalar $\theta$ and shear scalar $\sigma$ are decreasing function of time. They diverges to infinity as $t \to0$ and becomes zero at infinity. Moreover, as $ t \to \infty $, the ratio $\frac{\sigma}{\theta}\to0$. Hence, the investigated model approaches isotropy. Also, the deceleration parameter $q$ is found to be
\begin{equation}\label{Eq:24}
	q=\frac{d}{dt} \left(\frac{1}{H} \right)-1=-1+\frac{1}{m}  
\end{equation}

The sign of $q$ indicates whether the universe is accelerating or decelerating. A positive sign of $q$ implies a decelerating model, whereas a negative sign of $q$ indicates a accelerating model. According to recent cosmic studies, the expansion of the universe is rapidly accelerating. From equation \eqref{Eq:24}, it is clear that, the deceleration parameter $q$ is negative for $ m>1 $, thus the universe is accelerating.

Now, the expression of pressure $P$, energy density $\rho $ and magnetic flux $h^{2} $ for power law volumetric expansion model are obtained as follows:
\begin{equation} \label{Eq:25} 
	P=-\frac{\frac{3}{2} {c_{1}}^2 (\mu +2 \pi ) t^{2-8 m}+3 m (\mu  (4 m-3)+8 \pi  (2 m-1))}{8 \left(8 \pi ^2+7 \pi  \mu +2 \mu ^2\right) t^2}  
\end{equation}
\begin{equation} \label{Eq:26} 
	\rho =\frac{3 \left(-c_{1}^2\pi t^{2-8 m}+16 (\mu +\pi ) m^2-4 \mu  m\right)}{8 \left(8 \pi ^2+7 \pi  \mu +2 \mu ^2\right) t^2}
\end{equation}
\begin{equation} \label{Eq:27} 
	h^{2} =0
\end{equation}

\begin{figure}[h]
	\centering
	\includegraphics[width=0.75\linewidth]{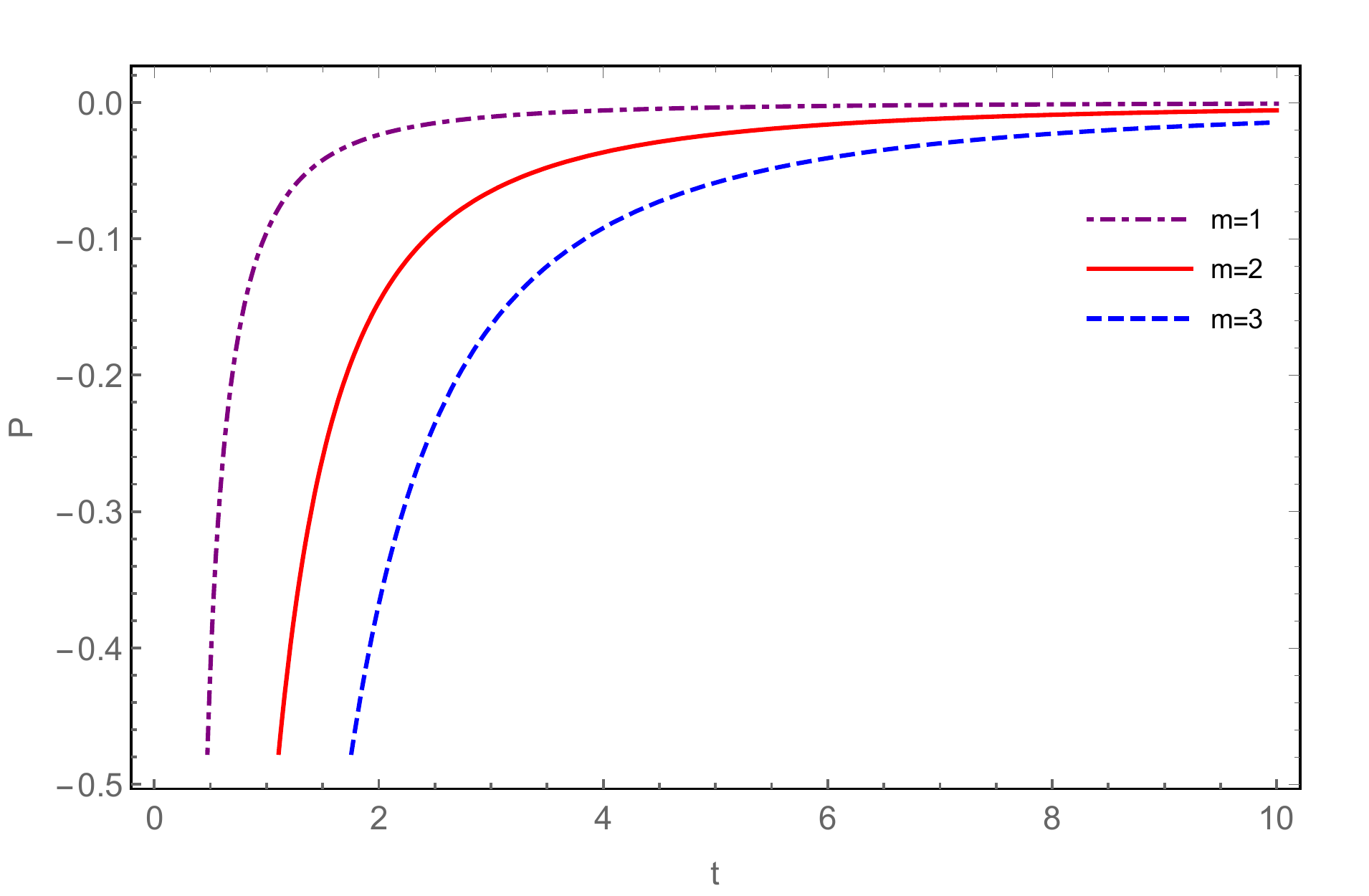}
	\caption{Plot of pressure $ (P) $ \textit{vs} time $t$ for $ c_{1}=1 $, $ \mu=0.1 $.}
	\label{fig:1}
\end{figure}
\begin{figure}[h]
	\centering
	\includegraphics[width=0.75\linewidth]{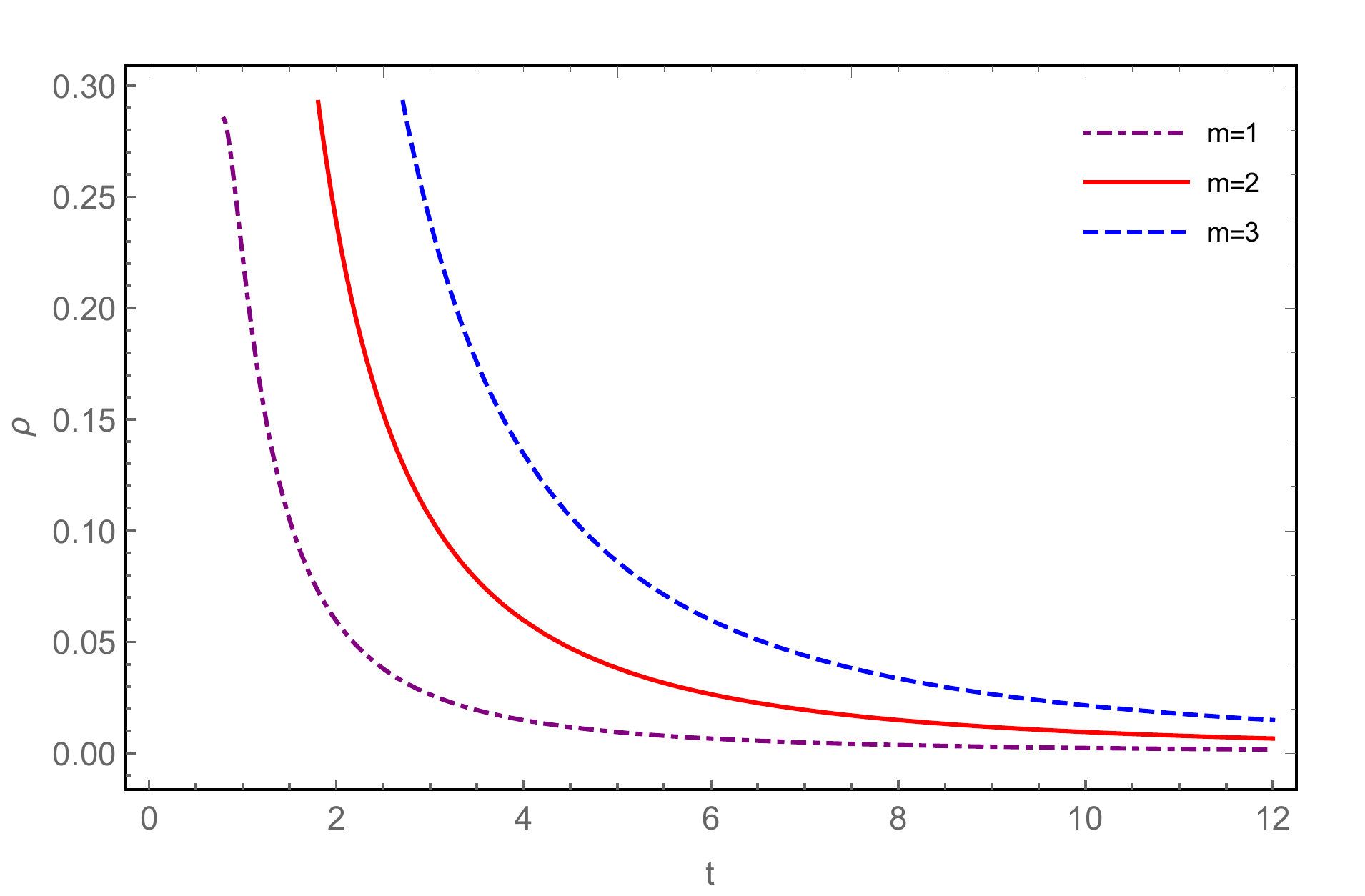}
	\caption{Plot of density $(\rho)$ \textit{vs} time $t$ for $ c_{1}=1 $, $ \mu=0.1 $.}
	\label{fig:2}
\end{figure}
The behaviour of pressure $ P $ and energy density $ \rho $ for the power law model is graphically depicted in figure \ref{fig:1} and \ref{fig:2} respectively. It is observe that the pressure is an increasing function of time $  t $; it is very small near $t=0$ and vanishes at infinite time $ t $. It is worth noting that the pressure is negative for different values of parameter $ m $, indicating that the SQM behaves like dark energy. Also, the energy density decreases as time $ t $ increases and it is infinte at $t=0$ and vanishes as $ t \to \infty $. For small values of $ m $, the density curve slowly approaches zero with increasing time. As parameter $ m $ becomes larger, the curve tends to a constant value.
\begin{figure}
	\centering
	\includegraphics[width=0.75\linewidth]{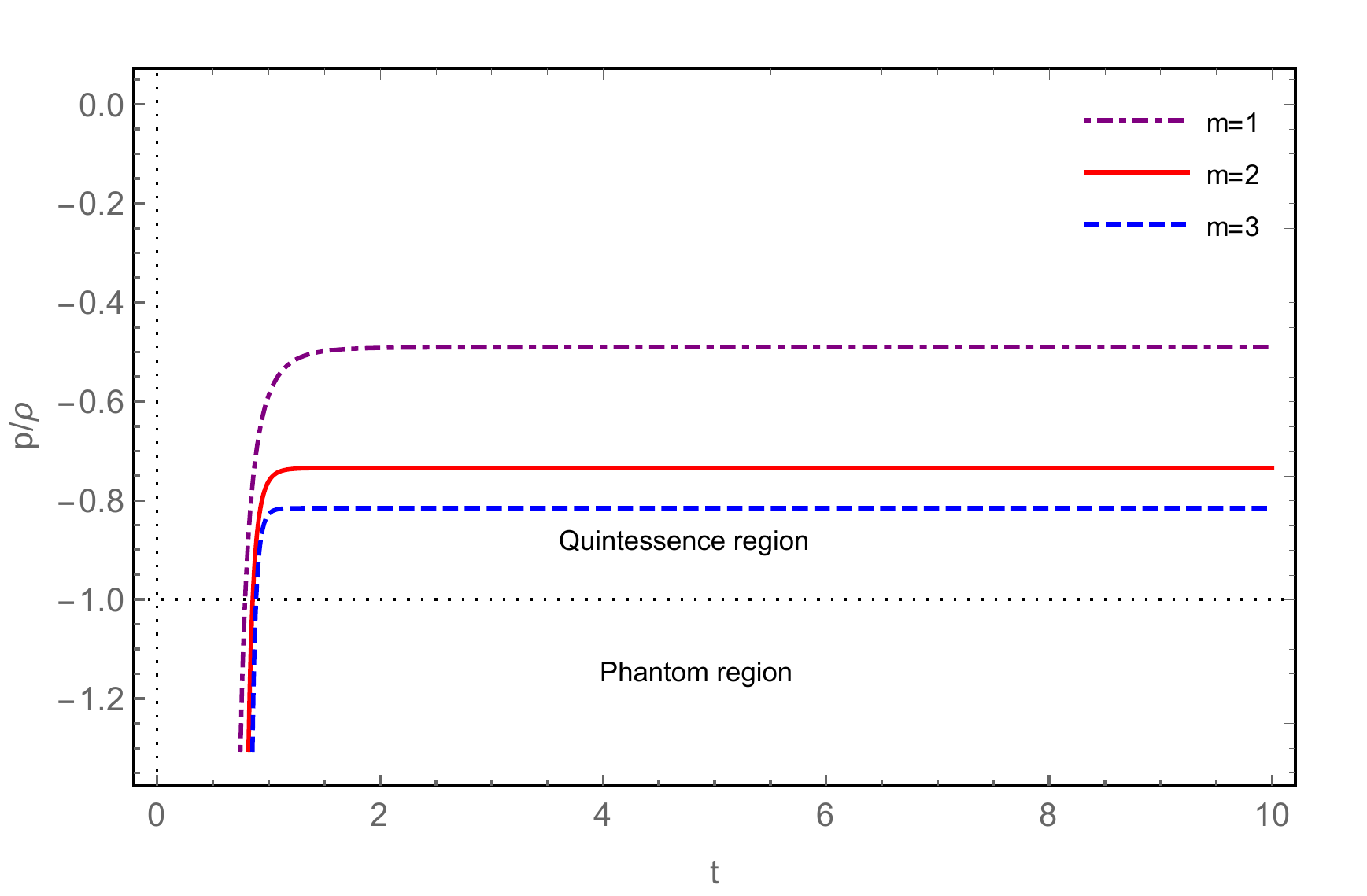}
	\caption{The plot of ratio ${p}/{\rho} $ \textit{vs} time $ t $ for $ c_{1}=1 $, $ \mu=0.1 $.}
	\label{fig:3}
\end{figure}
In the power law model, the graph of ${p}/{\rho} $ shows the dynamical behaviour with increasing time $ t $ as depicted in figure \ref{fig:3}.  It is observed that, initially, it evolves in the phantom region $ {p}/{\rho} < -1 $, crosses the phantom divide line $ {p}/{\rho} = -1 $, suddenly enters into the quintessence region $-1<{p}/{\rho}<-1/3$ and remains constants with increasing time. Suzuki et al. \cite{Suzuki et al. 2012} combined all four probes, such as SNe Ia with BAO, CMB, and $ H_{0} $ measurements, to determine the equation of state (EoS) parameter of dark energy (DE), which is $ \omega = -1.013^{+ 0.068}_{-0.073} $ for a flat $ w $CDM model. In the present case, for $ m>1 $, the value of ${p}/{\rho} $ is in the range $-1.38 \le \omega \le -0.89 $, which is consistent with the observational value of EoS ($ \omega $) of DE obtained by \cite{Suzuki et al. 2012,Ade et al. 2014}.  

\section{Exponential volumetric expansion model:}
For exponential volumetric expansion law, we assume the volume factor as
\begin{equation}\label{Eq:28}
	V=e^{4nt} 
\end{equation}
where $n$ is constant.

For this model, the values of $A$ and $B$ using the equation \eqref{Eq:28} are obtained as
\begin{equation}\label{Eq:29} 
	A=c_{2}^{1/4}exp\left[nt-\frac{c_{1}e^{-4nt}}{16n}\right]
\end{equation}
\begin{equation} \label{Eq:30} 
	B=c_{2}^{-3/4}exp\left[nt+\frac{3c_{1}e^{-4nt}}{16n}\right]
\end{equation}

$A$ and $B$ are constant at $ t=0 $, therefore the model has no singularity at $ t=0 $. Also, as $ t \to \infty $, both $A$ and $B$ tends to infinity. Now, the various cosmological parameters such as mean Hubble parameter $ H $, Expansion scalar $\theta $ and Shear scalar $\sigma $ are found to be
\begin{equation} \label{Eq:31} 
	H =n  
\end{equation}
\begin{equation} \label{Eq:32} 
	\theta =4n  
\end{equation}
\begin{equation} \label{Eq:33} 
	\sigma ^{2} =\frac{3}{2}\frac{c_{1}^2 }{e^{8nt}}
\end{equation}

From the equations \eqref{Eq:31}, \eqref{Eq:32} and \eqref{Eq:33}, it is observed that the Hubble parameter $ (H),$ exapansin scalar $(\theta)$ are constant i.e. the rate of expansion of the universe is constant. Shear scalar $(\sigma)$ is decreasing function of time. Therefore, $\frac{\sigma}{\theta}\to0$ as $ t \to \infty $ i.e. the model approach to isotropy.
Also, the deceleration parameter $q$ is obtain as
\begin{equation}\label{Eq:34}
	q=-1  
\end{equation}

The sign of deceleration parameter is negative, therefore the universe is accelerating.
The quantities such as the pressure $P$, energy density $\rho$ and magnetic flux $h^{2}$ for exponential volumetric expansion are obtained as
\begin{equation} \label{Eq:35} 
	P=-\frac{\frac{3}{2}c_{1}^2 (2 \pi +\mu) e^{-8 n t}+12 (4 \pi +\mu) n^2}{8 \left(8 \pi ^2+7 \pi  \mu +2 \mu ^2\right)}  
\end{equation}
\begin{equation} \label{Eq:36}
	\rho=\frac{e^{-8nt}\left(48(\pi+\mu)n^2 e^{8nt}-3\pi c_{1}^2\right)}{8\left(8\pi^2+7\pi\mu+2\mu^2\right)}
\end{equation}
\begin{equation} \label{Eq:37} 
	h^{2} =0 
\end{equation}

\begin{figure}[h]
	\centering
	\includegraphics[width=0.75\linewidth]{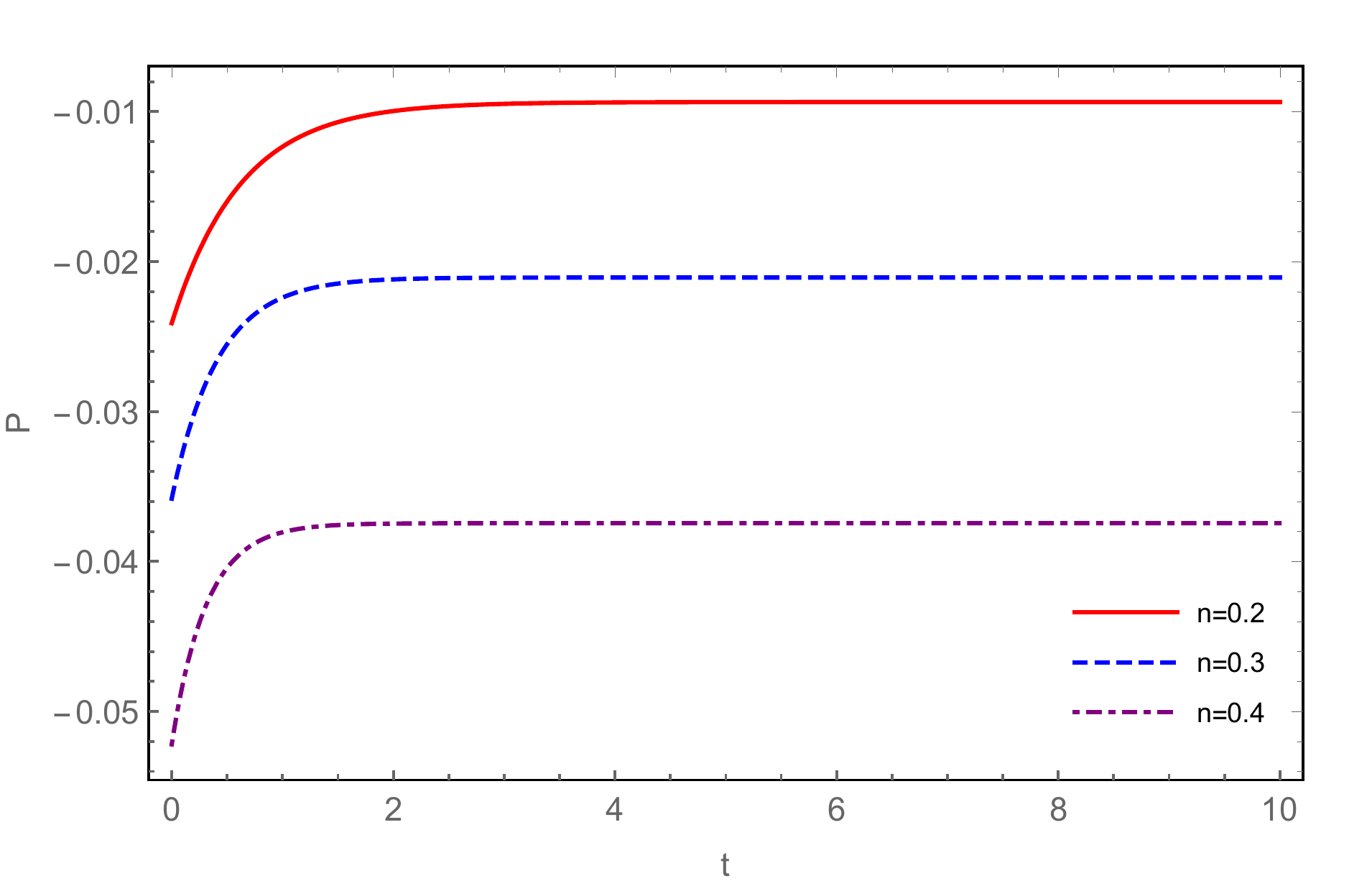}
	\caption{Plot of pressure $ (P) $  \textit{vs} time $t$ for $ c_{1}=1 $, $ \mu=0.1 $.}
	\label{fig:4}
\end{figure}
\begin{figure}[h]
	\centering
	\includegraphics[width=0.7\linewidth]{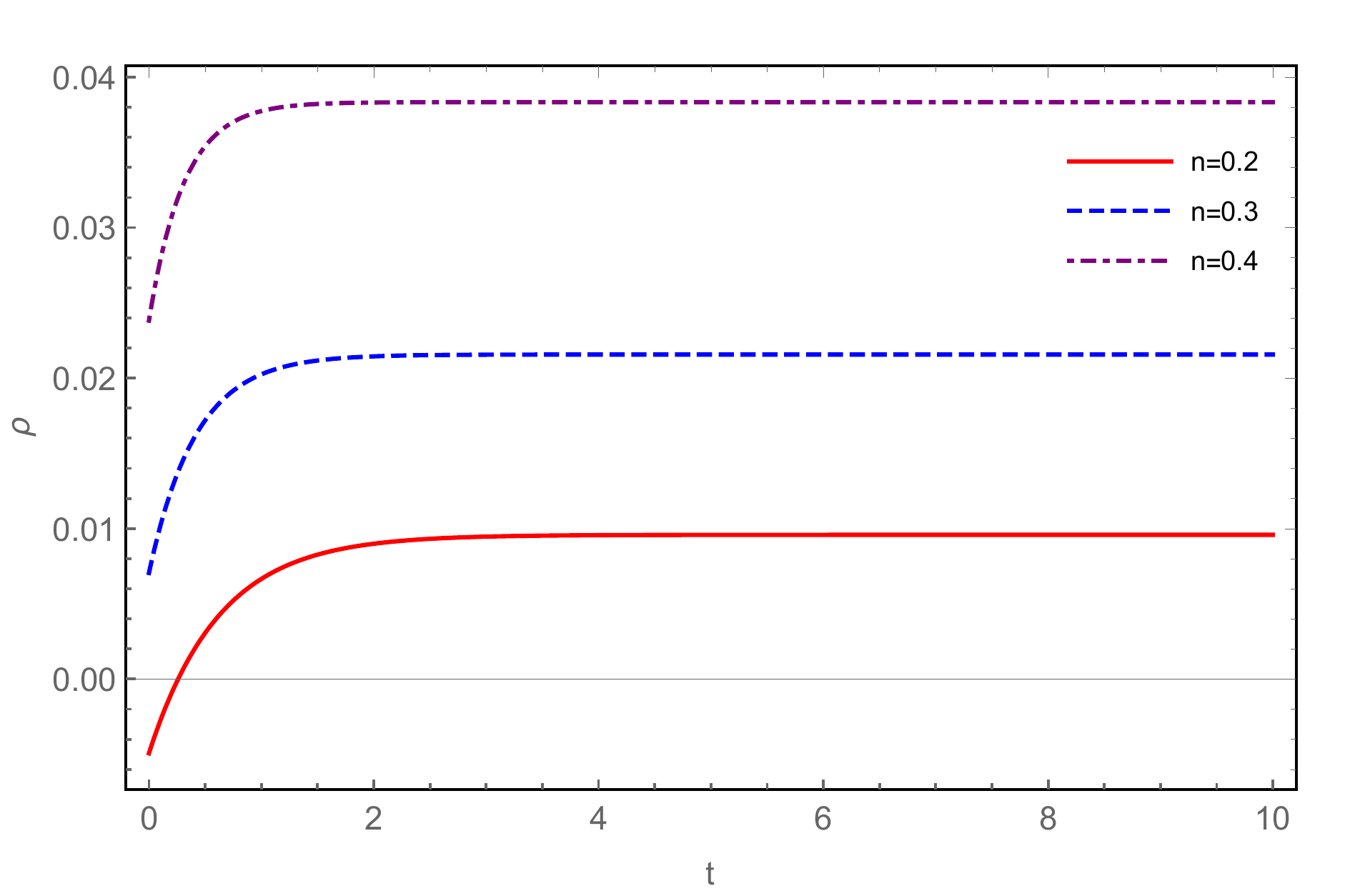}
	\caption{Plot of density $(\rho)$ \textit{vs} time $t$ for $c_{1}=1 $, $\mu=0.1$.}
	\label{fig:5}
\end{figure}
The graphical behaviour of pressure $ (P) $ and energy density $(\rho)$ \textit{vs} time $t$ is depicted in figure \ref{fig:4} and \ref{fig:5}. It is observe that, the energy density $\rho$ is decreasing function time $ t $. Initially energy density was constant near $ t=0 $ and slightly increased to its maximum value and remain constant with increasing time. The pressure is negative which reveals that the matter behave like dark energy.\\
\begin{figure}
	\centering
	\includegraphics[width=0.7\linewidth]{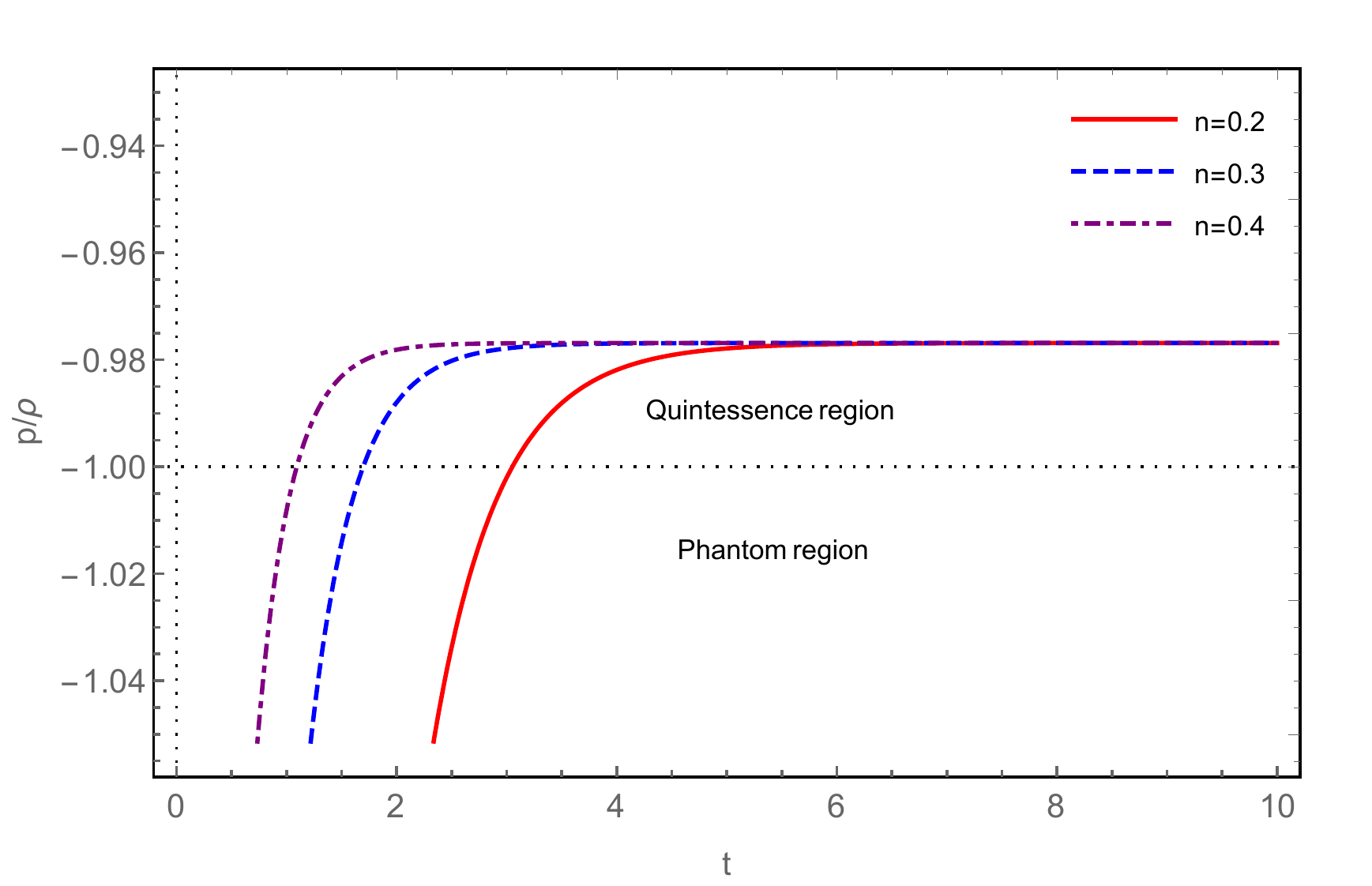}
	\caption{The plot of ratio $ {p}/{\rho} $ \textit{vs} time $ t $ for $ c_{1}=1 $, $ \mu=0.1 $.}
	\label{fig:6}
\end{figure}
In this case, the figure \ref{fig:6} represents the behaviour of $p/\rho $ with time $ t $. We observed that, for different value of $ n $, it begins to evolve in the phantom region and suddenly enters into the quintessence region with increasing time $ t $. Recently, Scolnic et al. \cite{Scolnic et al. 2018} combine the Planck 2015 CMB and SNe Ia measurements to calculate the best fit value for the EoS parameter $\omega= -1.026 \pm 0.041$. We observe that the value of $p/\rho $ is in the range $ -1.162 \le \omega \le -0.983$, which is consistent with the observed vale of EoS of DE obtained by \cite{Hinshaw et al. 2013,Scolnic et al. 2018}.

\section{Energy Conditions:}
The energy conditions are essential for understanding the geometry of universe. The Raychaudhuri's equation is used to obtain the energy conditions, which are a set of linear pressure-density combinations that decribe the energy density can never be negative and gravity attracts always. These energy conditions are defined and stated as follows:
\begin{itemize}
	\singlespacing
	\item SEC $\Rightarrow $ $\rho +3P \ge 0$,
	\item NEC $\Rightarrow $ $\rho+P \ge 0$,
	\item WEC $\Rightarrow $ $\rho\ge 0$, $\rho+P \ge 0$ and
	\item DEC $\Rightarrow $ $\rho>\left|P \right|\ge 0$.
\end{itemize}
Therefore, for first power law volumetric expansion model, the expression for NEC, SEC and DEC are obtained by using the equations \eqref{Eq:25} and \eqref{Eq:26} as
\begin{equation}
	\rho+P  =\frac{t^{-8 m-2} \left(6 m (\mu  (12 m-1)+8 \pi ) t^{8 m}-3c_{1}^2 (4 \pi +\mu ) t^2\right)}{16 \left(8 \pi ^2+7 \pi  \mu +2 \mu ^2\right)}
\end{equation}
\begin{equation}
	\rho+3P =-\frac{3 t^{-8 m-2} \left(c_{1}^2(3 \mu +8 \pi ) t^2+2 m (8 \pi  (4 m-3)-\mu  (4 m+5)) t^{8 m}\right)}{16 \left(8 \pi ^2+7 \pi  \mu +2 \mu ^2\right)}
\end{equation}
\begin{equation}
	\rho-P =\frac{3 \left(c_{1}^2\mu  t^{2-8 m}+2 m (\mu  (20 m-7)+8 \pi  (4 m-1))\right)}{16 \left(8 \pi ^2+7 \pi  \mu +2 \mu ^2\right) t^2}
\end{equation}

\begin{figure}[h]
	\centering
	\includegraphics[width=0.7\linewidth]{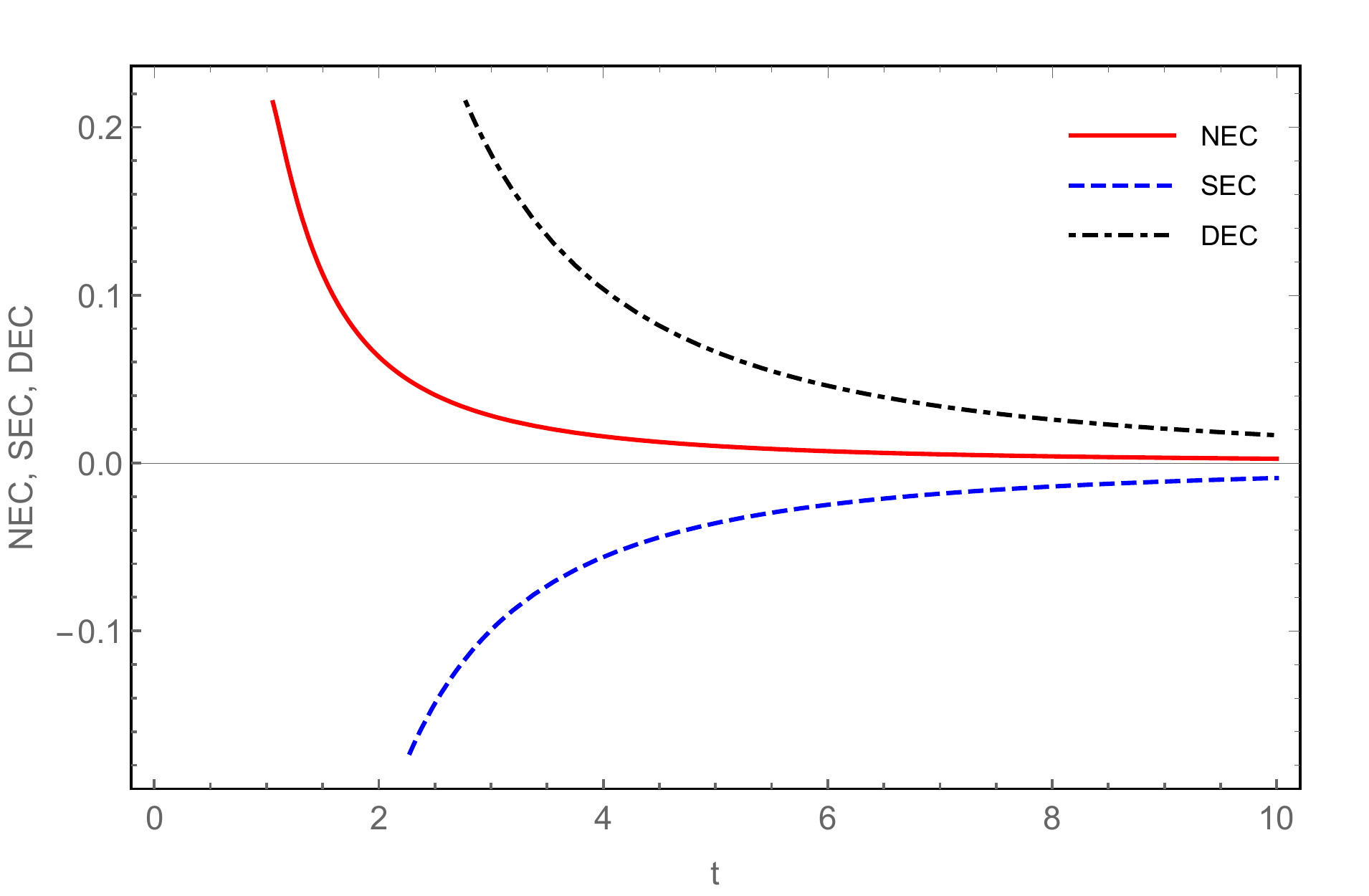}
	\caption{Plot of NEC, SEC and DEC \textit{vs} time $t$ for $ m=2 $, $ c_{1}=1 $, $ \mu=0.1 $.}
	\label{fig:7}
\end{figure}

\noindent Also, for Exponential volumetric expansion model, using equations \eqref{Eq:35} and \eqref{Eq:36}, NEC, SEC and DEC are obtained as
\begin{equation}
	\rho+P=\frac{e^{-8 n t} \left(72 \mu  n^2 e^{8 n t}-3 c_{1}^2(\mu +4 \pi )\right)}{16 \left(8 \pi ^2+7 \pi  \mu +2 \mu ^2\right)}
\end{equation}
\begin{equation}
	\rho+3P=-\frac{3 e^{-8 n t} \left(c_{1}^2(3 \mu +8 \pi )+8 (8 \pi -\mu ) n^2 e^{8 n t}\right)}{16 \left(8 \pi ^2+7 \pi  \mu +2 \mu ^2\right)}
\end{equation}
\begin{equation}
	\rho-P=-\frac{3 \left(c_{1}^2\mu  e^{-8 n t}+8 (5 \mu +8 \pi ) n^2\right)}{16 \left(8 \pi ^2+7 \pi  \mu +2 \mu ^2\right)}
\end{equation}
\begin{figure}[h]
	\centering
	\includegraphics[width=0.7\linewidth]{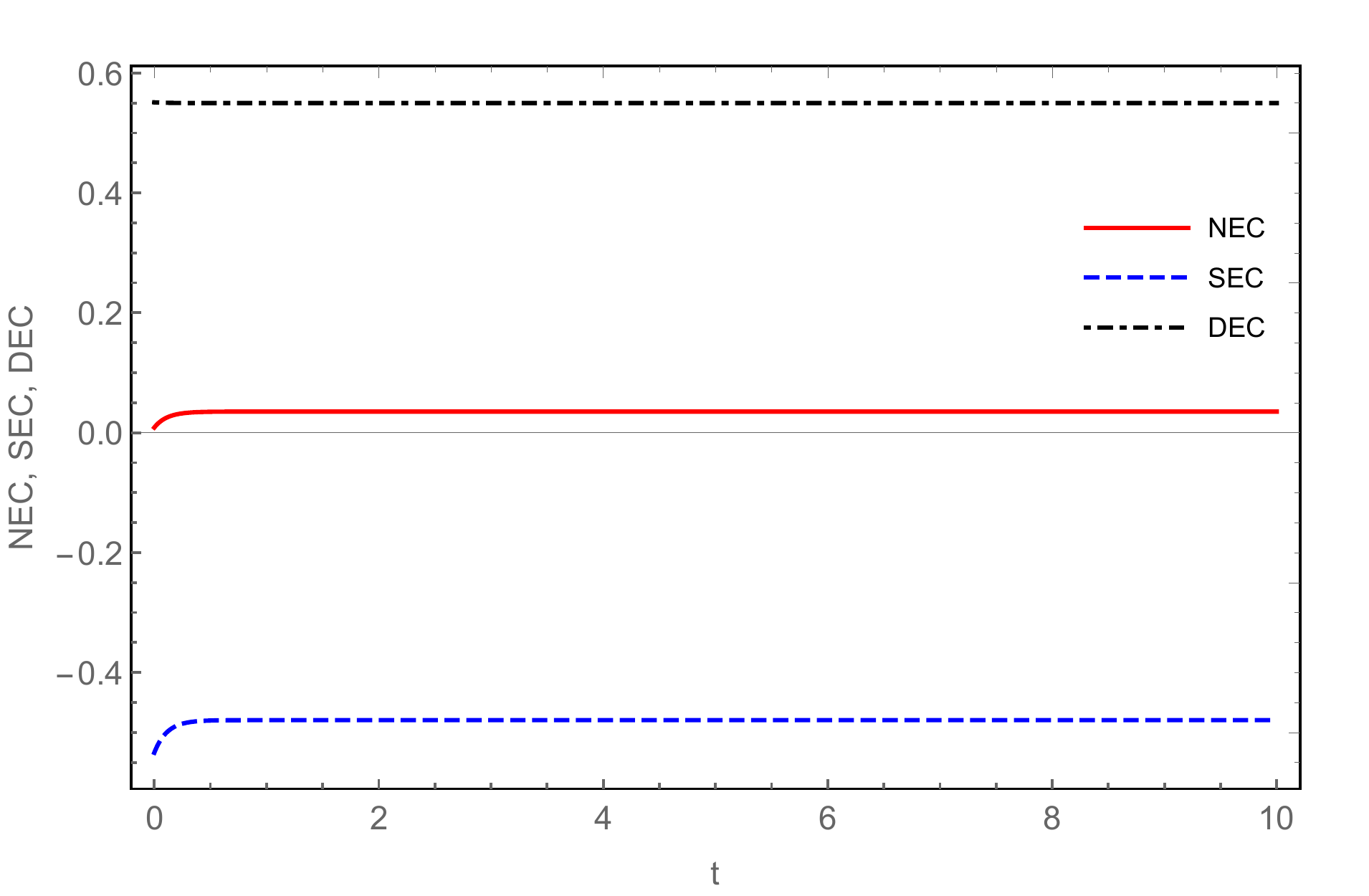}
	\caption{Plot of NEC, SEC and DEC \textit{vs} time $t$ for $ n=1 $, $ c_{1}=1 $, $ \mu=0.6 $.}
	\label{fig:8}
\end{figure}

Figures \ref{fig:7} and \ref{fig:8} show the graphical representation of NEC, SEC and DEC \textit{vs} time $t$  for power law and exponential law, respectively. From this, it is clear that the null energy condition (NEC) and dominant energy condition (DEC) are satisfied by both models, but the strong energy condition (SEC) is violated. Sahoo et al. \cite{Sahoo et al. 2016} has studied the energy condition and shows that only WEC and DEC satisfies for volumetric expansion model in $ f(R,T) $ gravity. Also, Alvarenga et al. \cite{Alvarenga et al. 2013} shown that energy conditions are  satisfied for suitable choice of inpute parameter in $ f(R,T) $ gravity theory.  Recently, Sahoo et al. \cite{Sahoo et al. 2021} has observed that, NEC and DEC are satisfy but SEC is violated. It is important to note that, our invstigation are more relevant than \cite{Sahoo et al. 2016, Alvarenga et al. 2013} and resemble with the investigation of \cite{Sahoo et al. 2021}, because both models satisfies all energy condtions except SEC.

\section{Conclusions:}

\noindent In this paper, we have studied Kaluza-Klein cosmological models with magnetized strange quark matter (MSQM) in $f(R,T)$ theory of gravity using power law and exponential volumetric expansions.
\begin{itemize}
	\item In power law expansion model, the shear and expansion scalars, as well as the Hubble parameter H, decrease over time.  The rate of expansion of the universe is very high near $  t=0 $. They diverge to infinity at $ t = 0 $ and vanish at infinite time. The energy density begins to decrease from a positive value, the pressure starts to increase from a negative value, and both approach zero at infinite time. It has been observed that the universe accelerates when $ m>1 $.
\end{itemize}
\begin{itemize}
	\item In the exponential volumetric expansion model, it is observed that the pressure and the energy density of the universe gradually increase to their maximums and then remain constant with increasing time. For $ n>0 $, the universe's expansion rate remains constant. Also, the negative value of the deceleration parameter $ q $ indicates that the universe is accelerating, which is consistent with recent findings.
\end{itemize}
\begin{itemize}
	\item It is important to note that, for both models, the negative pressure indicates that the matter behaves like dark energy, which causes the accelerated expansion of the universe. Moreover, the trajectory of $ p/\rho $ is lie in the range $-1.38 \le \omega \le -0.89 $ in the case first and $ -1.162 \le \omega \le -0.983$ in the second. The bounds of EoS parameter obtained by \cite{Suzuki et al. 2012} is $ \omega = -1.013^{+ 0.068}_{-0.073} $ and \cite{Scolnic et al. 2018} is $\omega= -1.026 \pm 0.041$. It seems that the obtained values of $ p/\rho $ in both cases are close to the observational bounds. Further, Aditya et al. \cite{Aditya et al. 2019} obtained the bounds of the EoS parameter in Lyra geometry as $ -1.3 \le \omega \le -0.8$. In Saez-Ballester theory, the bounds of the EoS parameter for the Kaluza-Klein model is $ -1.6 \le \omega \le 0.2$ \cite{Naidu et al. 2021}. When the values of the EoS parameter in scalar-tensor theories for the Kaluza-Klein model are compared to the range of $ p/\rho $ in the present model, the resultant value of $ p/\rho $ in $ f(R,T) $ gravity is found to be more appropriate than that obtained for the EoS parameter in scalar-tensor theories  \cite{Aditya et al. 2019,Naidu et al. 2021}. Also, the magnetic flux vanishes for both models and It's worth noting that our findings are consistent with those of \cite{Aktas 2017, Aktas and Aygun 2017, Khalafi and Malekolkalami 2021, Ozdemir and Aktas 2020}. Moreover, both the power law and exponential law models satisfy the NEC and DEC, but the SEC is violated throughout the evolution. 
\end{itemize}
\textbf{Acknowledgements:} The authors are greatful to the anonymous referee for constructive comments to upgrade the manuscript.

\end{document}